\documentclass[reprint,aps,prb]{revtex4-1}
\usepackage{color,graphicx,calc,url}
\usepackage{dcolumn}
\usepackage{bm}
\usepackage{amssymb}
\usepackage{amsmath}
\usepackage{wasysym}
\usepackage{mathrsfs} 
\usepackage{gensymb}
\usepackage{color}

\begin{document}
\title{Quantitative separation of the anisotropic magnetothermopower and planar Nernst effect by the rotation of an in-plane thermal gradient}
\author{Oliver Reimer$^{1*}$, Daniel Meier$^1$, Michel Bovender$^1$, Lars Helmich$^1$, Jan-Oliver Dreessen$^1$, Jan Krieft$^1$, Anatoly S. Shestakov$^2$, Christian H. Back$^2$,\\ Jan-Michael Schmalhorst$^1$, Andreas H\"utten$^1$, G\"unter~Reiss$^1$, and Timo~Kuschel$^{1,3}$\email{Electronic mail: oreimer@physik.uni-bielefeld.de}}
\affiliation{$^1$ Center for Spinelectronic Materials and Devices, Department of Physics, Bielefeld University, Universit\"atsstra\ss e 25, 33615 Bielefeld, Germany\\$^2$ Institute of Experimental and Applied Physics, University of Regensburg, Universit\"atsstra\ss e 31, 93040 Regensburg, Germany\\$^3$ Physics of Nanodevices, Zernike Institute for Advanced Materials, University of Groningen, Nijenborgh 4, 9747 AG Groningen, The Netherlands}

\date{\today}


\begin{abstract}
A thermal gradient as the driving force for spin currents plays a key role in spin caloritronics. In this field the spin Seebeck effect (SSE) is of major interest and was investigated in terms of in-plane thermal gradients inducing perpendicular spin currents (transverse SSE) and out-of-plane thermal gradients generating parallel spin currents (longitudinal SSE). Up to now all spincaloric experiments employ a spatially fixed thermal gradient. Thus anisotropic measurements with respect to well defined crystallographic directions were not possible. Here we introduce a new experiment that allows not only the in-plane rotation of the external magnetic field, but also the rotation of an in-plane thermal gradient controlled by optical temperature detection. As a consequence, the anisotropic magnetothermopower and the planar Nernst effect in a permalloy thin film can be measured simultaneously and reveal a phase shift, that allows the quantitative separation of the thermopower, the anisotropic magnetothermopower and the planar Nernst effect. 
\end{abstract}

\maketitle


Adding the spin degree of freedom to conventional charge-based electronics opens the field of spintronics \cite{Wolf:2001,Hoffmann:2015} with promising advantages such as decreased electric power consumption and increased integration densities. While spinelectronics use only voltages as driving force for currents, thermal gradients and the interaction between spins and heat currents have already been shown to provide new effects. Spin caloritronics investigate these interactions and promotes the search for applications such as heat sensors or waste heat recyclers \cite{Bauer:2012,Boona:2014}, that can improve thermoelectric devices. 
\\
One of the most important and well established phenomena in spin caloritronics is the longitudinal spin Seebeck effect (LSSE) \cite{Uchida:2010a,Uchida:2014,Uchida:2010b,Weiler:2012,Meier:2013a,Qu:2013,Kikkawa:2013}, which uses typically out-of-plane thermal gradients in magnetic thin films for the generation of a spin current parallel to the thermal gradient. This pure spin current is then injected into an adjacent non-magnetic conductor with high spin-orbit coupling, e.g. Pt, which transforms the spin current into an electric voltage via the inverse spin Hall effect (ISHE). In very recent investigations, the LSSE was even detected without any Pt and ISHE by the use of the anomalous Hall effect in Au \cite{Hou:2016} and by time-resolved magnetooptic Kerr effect in Au and Cu \cite{Kimmling:2016}.

Besides the application of an out-of-plane thermal gradient, effects driven by in-plane thermal gradients were also investigated. The transverse spin Seebeck effect (TSSE), the spin current generation perpendicular to an in-plane thermal gradient, was reported for metals \cite{Uchida:2008}, semiconductors \cite{Jaworski:2010} and insulators \cite{Uchida:2010c}. However, it has been noted that TSSE experiments in metals and semiconductors can be influenced by parasitic effects like the planar Nernst effect (PNE) \cite{Ky:1966b} or the anomalous Nernst effect (ANE) \cite{Nernst:1886}. The first occurs in samples with magnetic anisotropy \cite{Avery:2012}, while the latter can be attributed to unintended out-of-plane temperature gradients \cite{Huang:2011} due to heat flux into the surrounding region \cite{Schmid:2013} or through the electric contacts \cite{Meier:2013b}. Recently, the influence of inhomogeneous magnetic fields was added to the list of uncertainties for TSSE experiments \cite{Shestakov:2015}. Despite first reports, the TSSE could not be reproduced neither for metals \cite{Avery:2012,Huang:2011,Schmid:2013,Meier:2013b,Shestakov:2015,Bui:2014}, for semiconductors \cite{Soldatov:2014} nor insulators \cite{Meier:2015}.

Closely related to the recently reported spin Hall magnetoresistance (SMR) \cite{Nakayama:2013,Chen:2013,Vlietstra:2013,Althammer:2013} are magnetothermopower effects that were detected in bilayers of nonmagnetic conductors/ferromagnetic insulators \cite{Meier:2015,Meyer:2016}. In the first case an in-plane electric current is driven through a conductor with high spin-orbit coupling (e.g. Pt) deposited on a magnetic insulator. The interplay of the spin Hall effect (SHE) and the ISHE induces an anisotropic electric resistance in the conductor depending on the relative orientation between the spin polarization of the normal metal and the magnetization of the magnetic insulator. Whereas the SMR uses an electric potential to inject the charge current, the so-called spin Nernst magnetothermopower is driven by an in-plane thermal gradient and can be described by the recently discovered spin Nernst effect \cite{Meyer:2016} in combination with the ISHE. Hints of these effects were already observed as side-effects in experiments using heatable electric contact tips \cite{Meier:2015}. This is another example for the use of in-plane thermal gradients in spin caloritronics. However, the in-plane thermal gradients used so far are spatially fixed.\\

Different techniques to apply thermal gradients include Joule heating in an external heater \cite{Uchida:2008,Meier:2013a}, laser heating \cite{Walter:2011, Weiler:2012}, Peltier heating \cite{Uchida:2010a,Schmid:2013,Shestakov:2015}, current-induced heating in the sample \cite{Schreier:2013}, heating with electric contact needles \cite{Meier:2013b,Meier:2015} and on-chip heater devices\cite{Wu:2015}. In this work we present a setup, which uses Peltier heating as a method for heating and cooling purposes to cover a larger working temperature range. As a key feature, this setup allows the in-plane rotation of a thermal gradient $\nabla T$ and, thus, also the angle-dependent investigation of the anisotropic magnetothermopower (AMTP) and planar Nernst effect (PNE). The use of an infrared camera optically resolves the rotation of $\nabla T$.

In order to demonstrate the functionality of the setup, we will concentrate on the electric characterization of magnetothermopower effects, which can be measured under different angles of $\nabla T$. In analogy to the description of the anisotropic magnetoresistance, one can derive similar equations for magnetothermopower effects (supplementary informations (SI) chapter I). The setup of our experiment and the definition of the directions of $\nabla T$, and the external magnetic field $\vec{H}$ with respect to the coordinates are sketched in Fig. \ref{fig:sample} (a). When a temperature gradient $\nabla T$ is applied, its y-component $\nabla T_{\text{y}}$ will generate a longitudinal AMTP and thus an electric field $E_{\text{y}}$ in the y-direction. This longitudinal AMTP can be described by
\begin{align}
E_{\text{y}}&=-\left(S_+-S_-\cos2\varphi\right)\,|\nabla T| \, \sin{\varphi_{\text{T}}} \label{eq:amtp2}
\end{align}
with $S_+=\frac{S_{||}+S_{\perp}}{2}$, $S_-=\frac{S_{||}-S_{\perp}}{2}$ and $S_{||}$, $S_{\perp}$ beeing the Seebeck coefficients of the thermopower parallel and perpendicular to $\vec{M}$, respectively. $S_+$ originates from the ordinary, magnetic field independent thermovoltage whereas $S_-$ describes the magnetic field dependent part of the AMTP. $\varphi$ and $\varphi_\text{T}$ are the angles of the external magnetic field and $\nabla T$, respectively, with respect to the x-axis as defined in Fig. \ref{fig:sample}(a).
The transverse magnetothermopower also contributes to $E_\text{y}$ but is driven by $\nabla T_\text{x}$ and will be denoted as the PNE, which is determined by
\begin{align}
E_{\text{y}}&=-S_- \, \sin2 \varphi |\nabla T| \cos \varphi_{\text{T}} \label{eq:pne2} \ \ \ .
\end{align}
Summing up the AMTP and PNE contributions in the y-direction, we end up with
\begin{align}
E_{\text{y}}&=-(S_+ \, \sin \varphi_{\text{T}} + S_- \sin (2\, \varphi - \varphi_{\text{T}})) |\nabla T| \ \ \ . 
\end{align}
Thus, the angle $\varphi_{\text{T}}$ of the thermal gradient acts as a phase shift for the magnetization dependent part $S_-$ of the thermopower.

In Sec. I, the functionality of the setup is briefly explained and the measurement modes are introduced. In Sec. II, the setup is used to characterize the AMTP and the PNE in thin Ni$_{80}$Fe$_{20}$ (Py) films in dependence of the rotation angle $\varphi_{\text{T}}$ of $\nabla T$. Increasing $\varphi_{\text{T}}$ leads to a phase shift in $V_{\text{y}}$ and Eqs. (\ref{eq:amtp2}), (\ref{eq:pne2}) are used to split the superimposed voltage signals into the contributions of the AMTP and PNE. This enables a determination of the Seebeck coefficients parallel and perpendicular to the magnetization of the sample. 

\section{Experimental setup}

\begin{figure}
\centering
\includegraphics[width=3.4in]{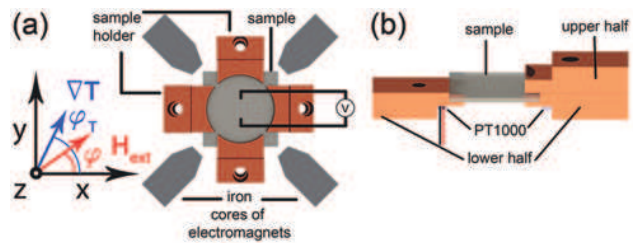}
\caption{(a) The sample is clamped between four circularly shaped holders, which can be heated independently. Thus, the variation of different applied $\nabla T_{\text{x}}$ and $\nabla T_{\text{y}}$ results in rotated net $\nabla T$. Two pairs of electromagnets rotated by $\pm$45$^\circ$ with respect to the x axis supply a rotatable in-plane magnetic field based on the superposition of the fields of both magnetic axes. (b) Each sample holder consists of a lower and upper half to reduce unintended out-of-plane thermal gradients in the sample. The temperatures are detected via PT1000 elements attached $\approx$ 2mm next to the sample.}
\label{fig:sample}
\end{figure}

The setup realizes an in-plane rotation of $\nabla T$ by four independently heated sample holders (Fig. \ref{fig:sample}(a)). The sample is clamped in the center of the sample holders and the application of different x and y temperature differences leads to a superpositioned net thermal gradient along $\varphi_{\text{T}}$. Four electromagnets arranged as shown in Fig. \ref{fig:sample} (a) additionally provide a rotatable in-plane magnetic field along $\varphi$. All electric measurements were conducted along the y axis of a sputter deposited Py thin film (5x5\, mm$^2$, 18\,nm thick) on MgO(001). To reduce parasitic effects due to symmetrically out-of-plane $\nabla T$, the heat is transferred into the sample using an upper and a lower half of the sample holder (Fig. \ref{fig:sample}(b)). This was already used in previous setups \cite{Meier:2013b,Meier:2015} and could successfully reduce unintended out-of-plane thermal gradients. PT1000 elements are glued at the backsides of each sample holder to detect the temperatures of the sample holders.

\begin{figure}[h]
\centering
\includegraphics[width=3in]{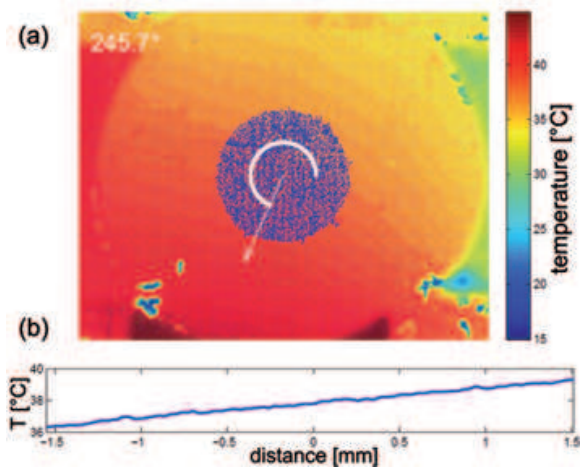}
\caption{(a) Thermographic picture of a Cu substrate with applied $\nabla T$ at $\varphi_{{\text{T}}} = 240^\circ$. The blue area represents the ROI, in which an averaged angle of 245.7$^\circ$ was calculated. (b) Temperature profile along $\varphi_{\text{T}} = 245.7^\circ$.}
\label{fig:cu}
\end{figure}

The successful rotation of $\nabla T$ is proven by an infrared camera, which clearly resolves the rotation of $\nabla T$ for different substrates and enables a quantitative analysis of the temperature profiles (see SI chapter II, with Ref. [\onlinecite{Thomson:1857,thompson:1975,lang:1992}]). Figure \ref{fig:cu} (a) shows a thermographic picture of a Cu substrate with $\nabla T$ applied along $\varphi_{\text{T}}=240^\circ$. After defining a Region of Interest (ROI, blue circle) the average angle of $\nabla T$ within the ROI can be calculated, symbolized by the white arc. Here, a deviation of the applied angle and the calculated angle of $\approx 6^\circ$ is detected. Taking a relative rotation between the setup and the camera by 2$^\circ$ into account, a mismatch of 4$^\circ$ is denoted as the uncertainty of $\varphi_{\text{T}}$. 

In the following, $V_{\text{y}}$ is averaged over five single measurements while the sample was kept at a base temperature of 308 K. When $V_{\text{y}}$ is measured as a function of the external magnetic field $\vec{H}$, which is varied from -150 Oe up to +150 Oe (black branch of results) and back down to -150 Oe (red branch of results), the measurement mode will be denoted as \emph{sweep measurement}.
When $V_{\text{y}}$ is measured in magnetic saturation as a function of $\varphi$, the \emph{field rotation measurement} mode was used. Here, the magnetization was kept saturated along the direction of $\vec{H}$ ($\Delta \varphi = \pm 3^\circ$) by using an external magnetic field of $200~\text{Oe}$, which then was rotated counterclockwise in the x-y plane (Fig. \ref{fig:sample}(a)).

\section{Results}
\subsection{$\Delta T$ dependence of the PNE}
\begin{figure}[h]
\includegraphics[width=3.4in]{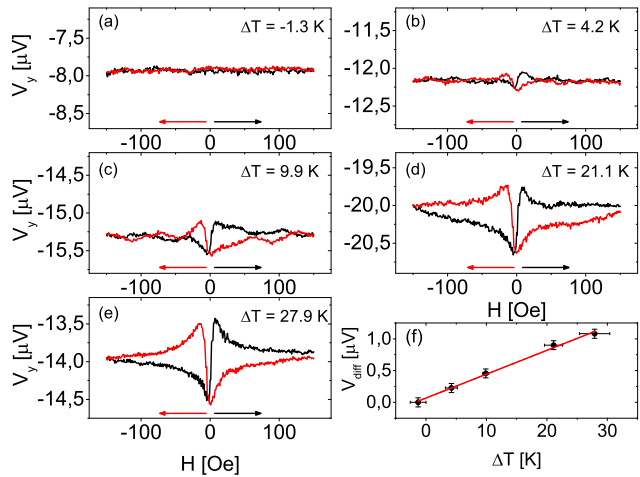}
\caption{(a) - (e) $V_{\text{y}}$ as a function of magnetic field for increasing $\Delta T$ and $\varphi=\varphi_{\text{T}}=0^\circ$. (f) $V_{\text{diff}}=V_{\text{max}}-V_{\text{min}}$ was calculated and averaged for each branch of each $\Delta T$ and plotted as a function of $\Delta T$, showing the expected linear dependence (Eq. (\ref{eq:pne2})).}
\label{fig:loopdT}
\end{figure}

Figure \ref{fig:loopdT} shows sweep measurements of $V_{\text{y}}$ with $\vec{H}$ aligned along $\varphi=0^\circ$. Here, $\Delta T$ was increased from $\approx\,0$ K to $\approx\,30\, K$ along $\varphi_{\text{T}}=0^\circ$. Keeping $\nabla T$ along the x direction and measuring the voltage only in the y direction excludes any AMTP contributions so that $V_{\text{y}}$ in Fig. \ref{fig:loopdT} only shows the PNE. In Fig. \ref{fig:loopdT}(a) a very low $\Delta T$ is applied along the x axis, which is too low to induce a detectable voltage along the y axis. Therefore, only the noise level ($\approx 50\,$nV) can be recorded. Depending on $\Delta T$, $V_{\text{y}}$ shows increasing peaks in the low magnetic field regime, and saturates for $|\vec{H}|\geq 140\, \text{Oe}$ (Fig. \ref{fig:loopdT}(a)-(e)). 

A similar experiment was conducted by Meier et al.\cite{Meier:2013b}, which is in good agreement with the data shown in Fig. \ref{fig:loopdT}. Slight deviations of the signal shape can be attributed to different magnetic anisotropies for different samples and small parasitic magnetic fields of the electromagnet due to the interaction of both magnetic axes (see SI chapter III). Starting with the increase of the magnetic field from -150 Oe to +150 Oe, for low negative field values the voltage of the PNE measurement (e.g. Fig. \ref{fig:loopdT}(e), black branch) first drops to a minimum voltage, lower than the saturation voltage, before it rises to a maximum value above the saturation voltage. Only then it decreases and saturates again. While decreasing the magnetic field after its maximum (red branch), again first the development of a minimum and then of a maximum is observed, before the voltage approaches the initial saturation value. 

For verifying the temperature dependence of the PNE, the peak-to-peak height in the low magnetic field regime is chosen as an indication of the PNE strength. The peak-to-peak height is quantified by $V_{\text{diff}}$, calculating the voltage difference between the maximum and minimum voltage for each branch and averaging them. Fig. \ref{fig:loopdT}(f) shows $V_{\text{diff}}$ vs. $\Delta T$. This correlation can be fitted linearly and therefore confirms the proportionality to $\Delta T$, as can be seen in Eq. (\ref{eq:pne2}).

\subsection{$\vec{H}$ angular dependence of the PNE}
\begin{figure}[h]
\includegraphics[width=3.4in]{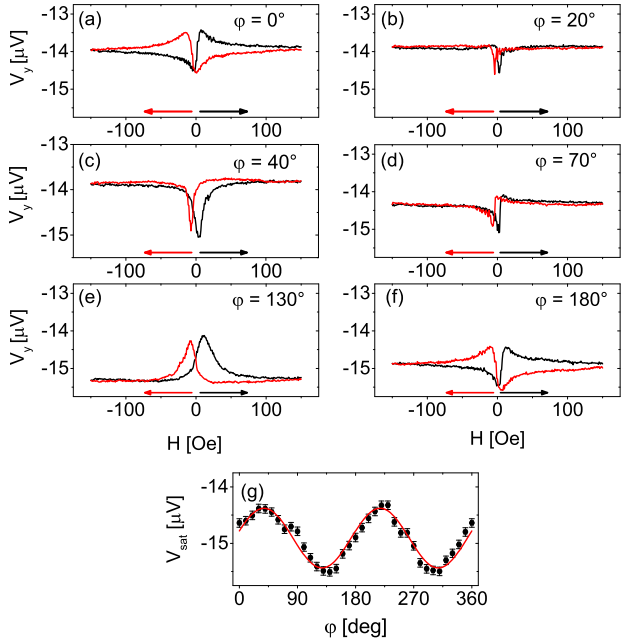}
\caption{(a) - (f) Measurement of $V_{\text{y}}$ against the magnetic field in a Py film on a MgO substrate. The temperature difference $\Delta T=30\text{ K}$ was kept constant along the x direction ($\varphi_{\text{T}}=0^\circ$). The in-plane angle $\varphi$ of the external magnetic field was varied. Data from $\varphi=0^\circ$ to $180^\circ$ are shown, since the $\sin{2\varphi}$ symmetry repeats the course for $\varphi\geq 180^\circ$. (g) The voltage $V_{\text{sat}}$ for each $\varphi$ was averaged in the range of $140 \text{ Oe}\leq|H|\leq 150$ Oe and plotted against $\varphi$, showing the theoretical predicted $\sin{2\varphi}$ dependence (Eq. (\ref{eq:pne2})).}
\label{fig:loop}
\end{figure}

Next, the sample was kept at a constant temperature difference of $\Delta T_{\text{x}}=30\text{ K}$, so the cold side was kept at $293\,$K and the hot side at $323\,$K. Sweep measurements were recorded for $0^\circ \leq \varphi \leq 360^\circ$ and six exemplary chosen curves in the range of $0^\circ \leq \varphi \leq 180^\circ$ are shown in Fig. \ref{fig:loop} (a)-(f). As before, $V_{\text{y}}$ saturates for high magnetic fields but shows differently shaped extrema, depending on $\varphi$. Fig. \ref{fig:loop} (a) shows the same data set as Fig. \ref{fig:loopdT} (e) with the appearance of a minimum and a maximum. Increasing $\varphi$ to 20$^\circ$ (Fig. \ref{fig:loop} (b)) changes the signal at the low magnetic regime into a minimum for both branches  with low intensity but similar shape. For $\varphi=40^\circ$ (Fig. \ref{fig:loop} (c)) the intensity of these minima increases until for $\varphi=70^\circ$ (Fig. \ref{fig:loop} (d)) the curves have changed their shape into a minimum and maximum again. But in contrast to Fig. \ref{fig:loop}(a) both branches have the same progression, thus, the magnetization reversal process is independent of the sweep direction of the magnetic field. For $\varphi=130^\circ$ (Fig. \ref{fig:loop} (e)) large, clearly separated maxima can be observed, which, in case of $\varphi=180^\circ$ (Fig. \ref{fig:loop} (f)), form a similar curve as for $\varphi=0^\circ$. For angles larger than $\varphi=180^\circ$ the curves from the range $0^\circ \geq \varphi \geq 180^\circ$ are repeated. 

\begin{figure}
\includegraphics[width=3.4in]{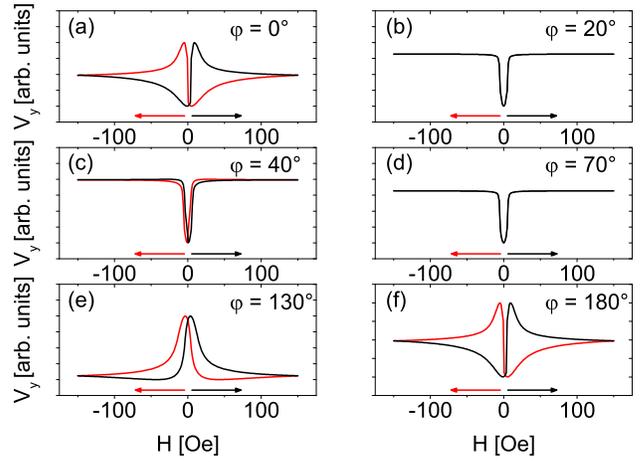}
\caption{Subsequent simulations based  on the Stoner-Wohlfarth model, described in SI chapter III, fit the experimental data of Fig. \ref{fig:loop}.}
\label{fig:simulation}
\end{figure}

The small signals of both branches for $\varphi=20^\circ, 70^\circ$ indicate magnetic easy axes in these directions \cite{Meier:2013b}. The appearance of two magnetic easy axes tilted by 50$^\circ$ can be explained by the non-parallel superposition of a uniaxial and a cubic magnetic anisotropy (see SI chapter III including Refs. [\onlinecite{Chen:1992,Park:1995, Daboo:1995, Kaibi:2015, Li:2015, Florczak:1991, Zhan:2007, Zhan:2009b, Zhan:2009, Kuschel:2011, Kuschel:2012}]).  
Furthermore, the experimental data can be fully understood and explained by simulations based on the Stoner-Wohlwarth model taking the geometry of the electromagnets into account (Fig. \ref{fig:simulation}, see SI chapter III including Refs. [\onlinecite{AMTEP2, AMTEP3, gurevich, UMA-CMA2, demag, CMAK, UMAK, mumax3}]). The signals for $\varphi$ = 20$^\circ$, 70$^\circ$ are the same in the simulations (Fig. \ref{fig:simulation} (b),(d)), whereas in the experiment they are not. Furthermore, the experiment observes a larger shift between the up and down trace, but beside of the mentioned issues, the simulations fit the experimental data qualitatively well.
\begin{figure}
\includegraphics[width=3in]{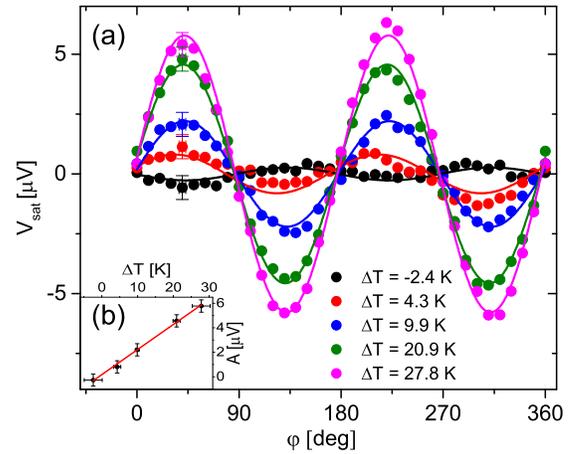}
\caption{(a) An external magnetic field of 200 Oe was rotated in-plane, keeping $\vec{M}$ saturated and aligned along $\varphi$. The measurement was repeated for increasing $\Delta T$ at $\varphi_{\text{T}}=0^\circ$ and fitted with $V_{\text{sat}}=y_0+A \, \sin{\left(2\varphi -\varphi_0\right)}$. The uncertainties $\delta \varphi$ and $\delta V_{\text{sat}}$ are only shown for the data points at $\varphi$ = 40$^\circ$ for reasons of better overview, (b) The fit parameter $A$ as a function of $\Delta T$ shows again the linear dependency with respect to $\Delta T$.}
\label{fig:rotation}
\end{figure}

Meier \emph{et al.} \cite{Meier:2013b} split the curves into a symmetric and antisymmetric part. A systematically observed antisymmetric part would indicate an ANE induced by an unintended out-of-plane $\nabla T$. Using this method for the data from Fig. \ref{fig:loop} does not show any systematic dependence of the antisymmetric contribution on the direction of the external magnetic field as it would be the case for the ANE. Therefore, we can exclude any unintended out-of-plane $\nabla T$ for the new setup, as we could for our other thermal setups \cite{Meier:2013b, Meier:2015}. The small non-systematic antisymmetric contributions can rather be explained by a non-perfect antisymmetric magnetization reversal process for some magnetic field directions due to an interplay of the magnetic anisotropy and field contributions mentioned in the SI chapter III.

Not only the shape of the curves but also the saturation voltage depends on $\varphi$. All saturation voltages for $|H|\geq 140$ Oe of each $\varphi$ were averaged, plotted vs. $\varphi$ and after subtraction of a linear temperature drift, $V_{\text{sat}}$ shows a clear sin2$\varphi$ dependence (Fig. \ref{fig:loop}(g)). $V_{\text{sat}}$ oscillates around an offset voltage of $\approx-15.0\, \mu$V, which originates from the ordinary thermovoltage, which is described in Eq. (\ref{eq:amtp2}) by $S_+$. Small deviations of $V_{\text{sat}}$ to the fit can be found around $\varphi=90^\circ,270°^\circ$, but an analysis of $V_{\text{sat}}-V_{\text{sin}2\varphi}$ reveals no systematical higher order measurement artefacts. Since the oscillation of Fig. \ref{fig:loop}(g) confirms the $\sin{2\varphi}$ dependence as predicted for the PNE by Eq. (\ref{eq:pne2}), further measurements in the rotation measurement mode for different $\Delta T$ are conducted to track down the PNE.

Five measurements were conducted for each $\Delta T$, averaged and plotted in Fig. \ref{fig:rotation} (a). All curves show the expected $\sin{2\varphi}$ oscillation so that based on Eq. (\ref{eq:pne2}) the data were fitted by $V_{{\text{sat}}}=y_0+A\,\sin{2(\varphi-\varphi_0)}$, very well confirming the agreement between the data and the theory of the PNE. Furthermore, plotting the amplitude $A$ of the fits vs. $\Delta T$ again shows the expected proportionality between the PNE and $\Delta T$ (Fig. \ref{fig:rotation} (b)).

\subsection{Phase shift of $\nabla T$ angular dependence of the AMTP and PNE}

\begin{figure}[h]
\includegraphics[width=3.4in]{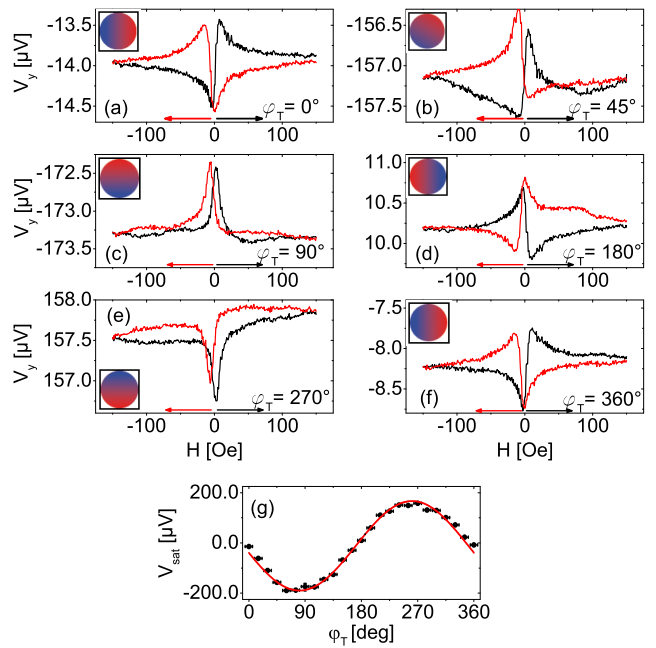}
\caption{(a) - (f) $V_{\text{y}}$ as a function of the magnetic field for $\varphi=0^\circ$ and various $\varphi_{\text{T}}$. The insets indicate the directions $\varphi_{\text{T}}$ of the thermal gradients. The sample was kept at a base temperature of 308 K with $\Delta T$=30 K. (g) The voltage $V_{\text{sat}}$ was averaged as described for Fig. \ref{fig:loop} (g) and plotted against $\varphi_\text{T}$. The data show a $\sin{\varphi_{\text{T}}}$ dependence attributed to $S_+$.}
\label{fig:looprotphiT}
\end{figure}

Next, the angle of the thermal gradient, $\varphi_{\text{T}}$, was continuously increased by 15$^\circ$ and sweep measurements were conducted for $\varphi =0 ^\circ$. Each curve again shows a saturation voltage for high magnetic fields and two extrema close to each other at around 0 Oe (Fig. \ref{fig:looprotphiT}(a)-(f)). In case of (a) the voltage measurement is carried out perpendicular to the thermal gradient, thus, the signal originates from the PNE ($E_{\text{y}}(\nabla T_{\text{x}}$)). In case of (c) the voltage measurement is conducted parallel to the thermal gradient because it was rotated to $\varphi_T=90^\circ$. Here, the voltage signal is attributed purely to the AMTP, since this effect needs a longitudinal $\nabla T \, (E_{\text{y}}(\nabla T_{\text{y}})$). 

\begin{figure}[h]
\includegraphics[width=3.4in]{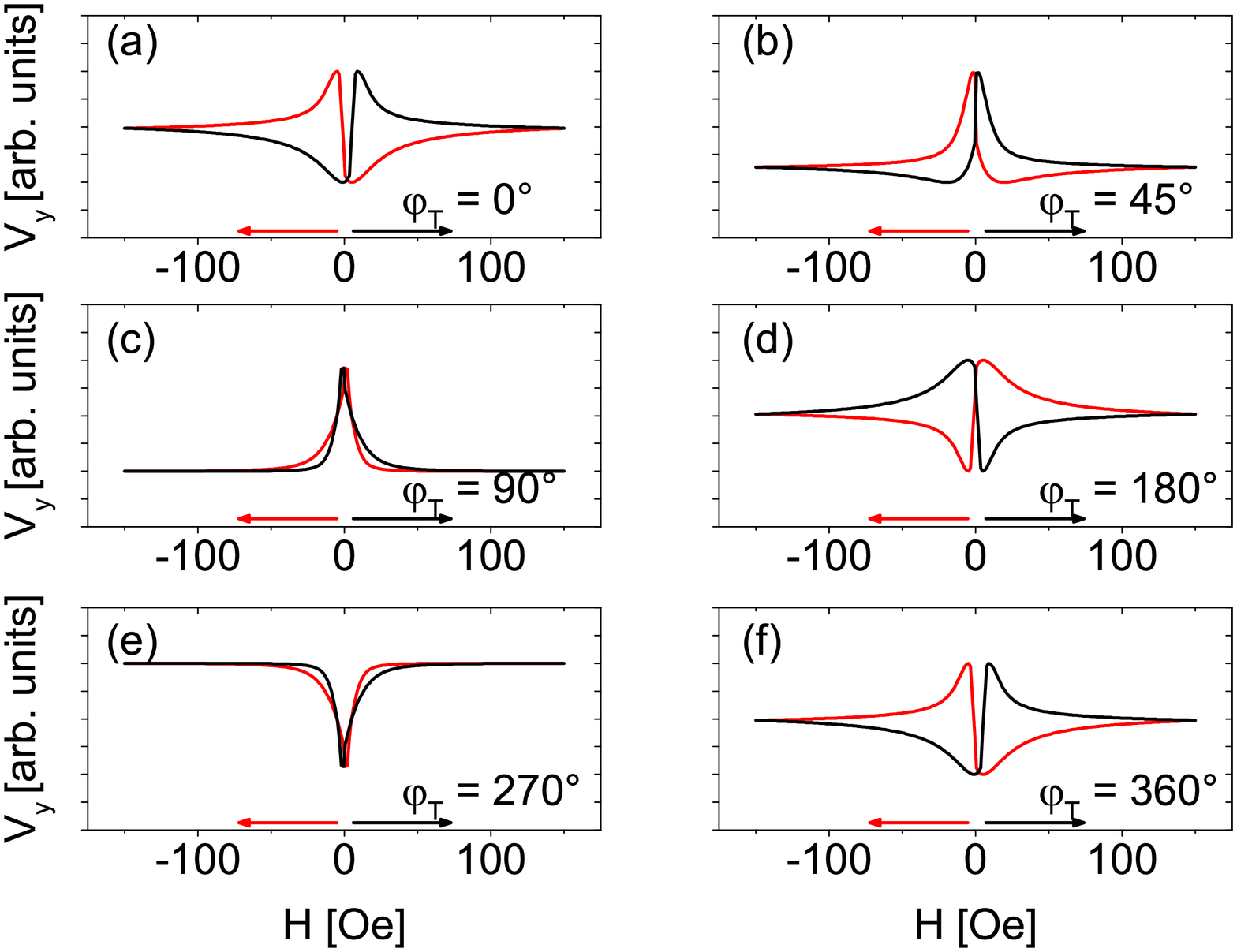}
\caption{The data of the sweep measurements for rotated $|\nabla T|$ (see  Fig. \ref{fig:looprotphiT}) can be simulated with the same model as used in Fig. \ref{fig:simulation}.}
\label{fig:simulation2}
\end{figure}

The results for $0^\circ<\varphi_{\text{T}}<90^\circ$ consist of a superposition of the PNE and the AMTP since for these $\varphi_{\text{T}}$, $\nabla T$ consists of a x and y component. This qualitative change in the signal can also be seen in the voltage features for low magnetic fields. Fig. \ref{fig:looprotphiT}(a) shows the same signal progression as described in section B, with the formation of a minimum before crossing 0 Oe. Increasing $\varphi_{\text{T}}$ now suppresses this minimum before the zero crossing point until for $\varphi_{\text{T}}=90^\circ$ only two sharp maxima are shaped. Due to the rotation of $\nabla T$ the relative orientation of $\vec{M}$ with respect to $\nabla T$ changes for different $\varphi_T$, thus, leading to changing contributions of the PNE and AMTP to the measured voltage signal. Again, the trace of the voltage signal can be fairly simulated as can be seen in Fig. \ref{fig:simulation2}.

Figure \ref{fig:looprotphiT}(g) shows the saturation voltages of Fig. \ref{fig:looprotphiT}(a)-(f) vs. $\varphi_{\text{T}}$. In contrast to Fig. \ref{fig:loop}(g), where the oscillation of $V_\text{sat}(\varphi)$ is only due to the PNE, Fig. \ref{fig:looprotphiT}(g) identifies the contribution of the ordinary, magnetic field independent Seebeck effect $V_{\text{sat}}(\varphi_T)$, expressed by $S_+$ in Eq. (\ref{eq:amtp2}). Since $V_{\text{y}}$ is measured, the rotation of $\nabla T$ leads to a $\sin{\varphi_{\text{T}}}$ shaped projection of $\nabla T$ on the y axis, resulting in a sine shaped $V_{\text{y}}$ signal. The nonmagnetic Seebeck signal is three orders of magnitude higher than the one of the PNE, while the magnetic field dependent part of the AMTP is expected to be of the same order of magnitude than the PNE. 

\begin{figure}
\includegraphics[width=3.4in]{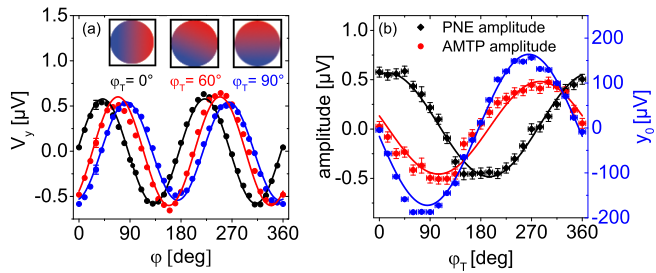}
\caption{(a) $V_{\text{y}}$ was measured in saturation (200 Oe) while rotating $\vec{H}_{\text{ext}}$ for different $\nabla T$ angles $\varphi_{\text{T}}$. The uncertainties $\delta \varphi$ and $\delta \, V_y$ are only shown for the data points at $\varphi$ = 30$^\circ$ for reasons of better overview. Again, the insets visualize the directions $\varphi_{\text{T}}$ of the thermal gradients. Increasing $\varphi_{\text{T}}$ results in a phase shift in the rotation measurement and further shifts the offset position $y_0$ from -4.83$\, \mu$V ($\varphi_{\text{T}}=0^\circ$) over \mbox{-187.6$\, \mu$V} ($\varphi_{\text{T}}=60^\circ$) to -187.1$\, \mu$V ($\varphi_{\text{T}}=90^\circ$). The phase shift indicates a superposition of PNE and AMTP. Therefore, the data were fitted with a $\cos 2\varphi$ (AMTP) and a $\sin 2\varphi$ (PNE) superposition. (b) The amplitudes of the $\cos 2\varphi$ and the $\sin 2\varphi$ contributions as well as the offset y$_0$ in the rotation measurement were plotted against $\varphi_{\text{T}}$ showing the expected cos- (PNE), sin- (AMTP) and sin- (ordinary Seebeck effect) dependence on $\varphi_{\text{T}}$.}
\label{fig:rotphiT}
\end{figure}

For the direct comparison of the different AMTP and PNE contributions, rotation measurements for $0^\circ \leq \varphi_{\text{T}} \leq 360^\circ$ were conducted. Fig. \ref{fig:rotphiT}(a) shows rotation measurements for three different $\varphi_{\text{T}}$, with offset voltages $y_0$ substracted. As described above, the oscillating signal of $V_{\text{y}}$ at $\varphi_{\text{T}}=0^\circ$ originates purely from the PNE and the oscillation of $\varphi_{\text{T}}=90^\circ$ purely from the AMTP. Since for all $\varphi_{\text{T}}$ in between we obtain a superimposed signal of both, the rotation measurements for all $\varphi_{\text{T}}$ were fitted with 
\begin{align}
V_{\text{y}}(\varphi,\,\varphi_{\text{T}})&=A(\varphi_{\text{T}})\, \sin{2(\varphi-\varphi_0)} \nonumber \\
&\ \ \ +B(\varphi_{\text{T}})\, \cos{2(\varphi-\varphi_0)}+y_{0}(\varphi_{\text{T}}) \ \ \ , \\ \nonumber \\ 
\text{with} \nonumber \\
 A(\varphi_{\text{T}}) &= \, - S_- \, |\nabla T|\, d  \, \cos\varphi_{\text{T}} \label{eq:fitA} \ \ \ , \\
 B(\varphi_{\text{T}}) &= \, S_- \, |\nabla T| \, d \, \sin\varphi_{\text{T}} \label{eq:fitB}\ \ \ , \\
 y_0(\varphi_{\text{T}})&=\, - S_+\, |\nabla T| \, d \, \sin\varphi_{\text{T}} \label{eq:fity0}
\end{align}
based on Eqs. (\ref{eq:amtp2}), (\ref{eq:pne2}). Here, the fit parameters $A$ and $B$ indicate the amplitudes of the PNE and AMTP, respectively. $d$ is the distance of the electric contacts and $y_0$ is the offset in $V_{\text{y}}$, which mirrors the ordinary Seebeck effect expressed as $S_+$. When $V_\text{y}$ is plotted vs. $\varphi$, Fig. \ref{fig:rotphiT}(a) shows the superposition of both effects, which leads to a phase shift of the measured signal for $\varphi_{\text{T}}>0^\circ$, described by Eq. (3). The $\sin{2\varphi}$ dependence (for $\varphi_{\text{T}}=0^\circ$), expected for the PNE (Eq. (\ref{eq:pne2})) is shifted to a $ -\cos{2\varphi}$ dependence (for $\varphi_{\text{T}}=90^\circ$), predicted for the AMTP (Eq. (\ref{eq:amtp2})).\\

In addition to the detected phase shift in the resulting signal, the change of the PNE (AMTP) ratio for each $\varphi_{\text{T}}$ can be revealed by plotting the fit amplitude $A$ ($B$) vs. $\varphi_{\text{T}}$ (Fig. \ref{fig:rotphiT}(b)). The result clearly shows a cosine (PNE) and a sine (AMTP) dependence of the amplitudes on $\varphi_{\text{T}}$ as determined by Eqs. (\ref{eq:fitA}), (\ref{eq:fitB}). The resulting cosine and sine fit functions result in a PNE amplitude of $(0.53\pm 0.05) \,\mu\text{V}$ and an AMTP amplitude of ($-0.47\pm 0.05)\,\mu\text{V}$. Within the measurement uncertainty the absolute value of the magnitudes of both effects are the same as it was expected from Eqs. (\ref{eq:fitA}), (\ref{eq:fitB}). Additional to the amplitudes, plotting $y_0$ vs. $\varphi_T$ gives a sine function as Eq. (\ref{eq:fity0}) predicts. 

With these findings we can determine the thermovoltages $S_-\, |\nabla T|\, d$ and $S_+\, |\nabla T|\, d$. 
Averaging the absolute values of the amplitudes of $A$ and $B$ results in 
\begin{align}
S_-\, |\nabla T|\, d=-(0.50 \pm 0.05) \mu\text{V} \label{eq:s-}
\end{align}
and the amplitude of $y_0$ gives 
\begin{align}
 S_+ \, | \nabla T|\, d=(168 \pm 4) \, \mu \text{V}\ \ \ . \label{eq:s+}
\end{align}
With the same $\nabla T$ and $d$, Eqn. (\ref{eq:s-}) and (\ref{eq:s+}) are used to calculate the relative change of the anisotropic Seebeck coefficient $\Delta S$ 
\begin{align}
\Delta S &= \frac{S_{||}-S_{\perp}}{S_{||}}\\
         &= \frac{2S_-}{S_+ + S_-}=-(0.60\pm 0.08)\% \ \ \ .
\end{align}
This calculation shows that the magnetothermopower perpendicular to the magnetization is $0.60\%$ stronger than parallel to the magnetization. The rotation of $|\nabla T|$ was used to sucessfully seperate PNE from AMTP measurements, which is observed by the subsequent shift of a sin- to a cos-dependence of the magnetic field rotation measurement. 

\section{Conclusion}
In conclusion, a novel setup was realized, which allows a well-defined rotation of an in-plane thermal gradient by superposition of two perpendicular thermal gradients of variable strength. Thus, the simultaneous measurement of the AMTP and PNE is made possible. The functionality of the setup was demonstrated and analyzed by an infrared camera and could further be verified by the subsequent electric analysis of magnetothermopower effects in a permalloy thin film on MgO(001). First, the proportionality dependency of the PNE to the temperature difference was shown. Second, a sweep of the external magnetic field was conducted for different angles and spatial fixed $\nabla T$, showing a repetition of the voltage signal for angles larger than 180$^\circ$. Plotting the saturation voltages vs. the magnetic field angle $\varphi$ shows a $\sin{2\varphi}$ dependency, verifying the theoretical predictions. By only rotating a high magnetic field, these $\sin{2 \varphi}$  oscillations can be measured directly. Measuring them for rotated $\nabla T$ leads to a phase shift until for $\varphi_T=90^\circ$ the $\sin {2 \varphi}$ oscillation of the magnetic field angular dependence is shifted to a $\cos{2\varphi}$ oscillation. This shift is due to a superposition of the PNE and AMTP and is the proof for a successful and controlled rotation of $\nabla T$. It further enables the splitting of the measured signal into $\varphi_T$ dependent contributions of the PNE, AMTP and ordinary Seebeck effect. This allows us to estimate the thermovoltages parallel and perpendicular to the magnetization using Eqs. (\ref{eq:s-}), (\ref{eq:s+}) to 
\begin{align}
V_{S,||}&=(S_++S_-)|\nabla T|\, d \nonumber \\
&= (168 \pm \, 5)\, \mu \text{V} \nonumber
\end{align}
and
\begin{align}
V_{S,\perp}&=(S_+-S_-)|\nabla T|\, d \nonumber\\
&=(169 \pm \, 5)\, \mu \text{V} \nonumber \ \ \ , 
\end{align}
resulting in a relative magnitude of the anisotropic magnetothermopower of $\Delta S=-(0.60\pm 0.08)\%$. 

After proving the rotation of $| \nabla T |$ with respect to the crystal structure, this setup is a promising tool to establish this method in future spin caloric experiments such as detailed anisotropy investigations of the spin Nernst magnetothermopower.

\begin{acknowledgments}
The authors gratefully acknowledge financial support by the Deutsche Forschungsgemeinschaft (DFG) within the priority program Spin Caloric Transport (SPP 1538).
\end{acknowledgments}

\section*{Author contributions}
O. R., M. B., and T. K. designed the experimental setup with the input of D. M., L. H., J.-O. D., J.-M. S., A. H., and G.R.; O. R. prepared and characterized the sample with the help of J. K. and performed the measurements; A. S. performed the theoretical simulations with the input of O. R. and T. K. in collaboration with C. B.;  O. R. and T. K. analyzed the data and wrote the manuscript with the input of all authors.

\section*{Additional information}
\textbf{Competing financial interests:} The author declare no competing financial interests. \\
\textbf{Correspondence:} All correspondence shall be directed to oreimer@physik.uni-bielefeld.de.


\end{document}


\title{- Supplementary Informations - \\
Quantitative separation of the anisotropic magnetothermopower and planar Nernst effect by the rotation of an in-plane thermal gradient}
\author{Oliver Reimer$^1$, Daniel Meier$^1$, Michel Bovender$^1$, Lars Helmich$^1$, Jan-Oliver Dreessen$^1$, Jan Krieft$^1$, Anatoly S. Shestakov$^2$, Christian H. Back$^2$,\\ Jan-Michael Schmalhorst$^1$, Andreas H\"utten$^1$, G\"unter~Reiss$^1$, and Timo~Kuschel$^{1,3}$\email{Electronic mail: oreimer@physik.uni-bielefeld.de}}
\affiliation{$^1$ Center for Spinelectronic Materials and Devices, Department of Physics, Bielefeld University, Universit\"atsstra\ss e 25, 33615 Bielefeld, Germany\\$^2$ Institute of Experimental and Applied Physics, University of Regensburg, Universit\"atsstra\ss e 31, 93040 Regensburg, Germany\\$^3$ Physics of Nanodevices, Zernike Institute for Advanced Materials, University of Groningen, Nijenborgh 4, 9747 AG Groningen, The Netherlands}

\begin{abstract}
In the first part of the supplementary informations we derive a theoretical description of the magnetothermopower effects and an equation optimized for our measurement geometry. The second part describes the details of the experimental setup and its calibration. It is followed by the optical investigation via an infrared camera of different substrates. In the last part, we further show that the electric measurements can be very well fitted by simulations based on the Stoner-Wohlfarth-model that are supported by x-ray diffraction measurements of the used sample.
\end{abstract}

\maketitle

\onecolumngrid

\section{Theoretical background of magnetothermal effects}
\begin{figure}[h]
\centering
\includegraphics[width=4in]{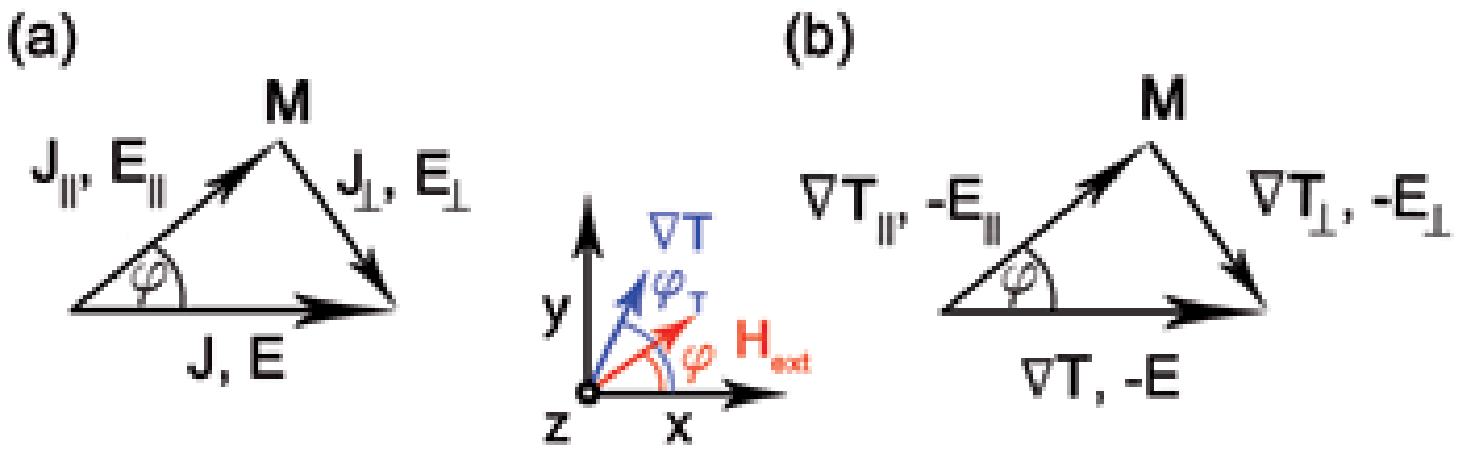}
\caption{(a) When an electric current is driven through a ferromagnetic conductor along x, the electric resistance parallel and perpendicular to M is different due to the AMR. The different electric fields parallel and perpendicular to M result in different electric fields along the x- and y-direction strongly depending on the angle between the electric current and M. (b) Thermal analogon to (a): A temperature gradient is the driving force of an electric current along x. Due to the anisotropic orbitals of the atoms, the parallel and perpendicular Seebeck coefficients differ from each other. Thus, the measured voltage is strongly dependent on the direction of the magnetization.}
\label{fig:dreieck}
\end{figure}

The anisotropic magnetoresistance (AMR) describes the difference of the electric resistivity parallel and perpendicular to the magnetization direction of a ferromagnetic conductor \cite{Thomson:1857}. The overlap of atom orbitals changes with the direction of the magnetization due to spin-orbit coupling. This affects the scattering cross section and, therefore, the electric resistivity, which is typically reduced for a magnetization direction perpendicular to the electric current. Thus, the measured voltage across a ferromagnet depends on the directions of current and magnetization. Assuming a current density along x, $|\vec{J}|=J_{\text{x}}$ (Fig. \ref{fig:dreieck} (a)), the longitudinal electric field $E_{\text{x}}$ depends on the electric field components parallel and perpendicular to the magnetization. Following Thompson \emph{et al.}, these components $E_{||}$ and $E_{\perp}$ can be described by J, the resistivity $\rho$ and the angle $\varphi$ between the magnetization and the x-axis\cite{thompson:1975}. One obtains
\begin{align}
	E_{||}&=\rho_{||} \, J_{\text{x}} \, \cos{\varphi} \\
	E_{\perp}&=\rho_{\perp} \, J_{\text{x}} \, \sin{\varphi} \ \ \ .
\end{align}
The projections of E$_{||}$ and E$_{\perp}$ onto the x-axis are
\begin{align}
E_{\text{x},||}&=E_{||}\, \cos{\varphi}= \rho_{||} \, J_{\text{x}} \, \cos^2{\varphi}\\
E_{\text{x},\perp}&=E_{\perp}\, \sin{\varphi}= \rho_{\perp} \, J_{\text{x}} \, \sin^2{\varphi}
\end{align}
leading to
\begin{align}
E_{\text{x}}&=E_{\text{x},||}+E_{\text{x},\perp} \\
&=\rho_{||}\, J_{\text{x}} \, \cos^2\varphi + \rho_{\perp} \, J_{\text{x}} \, \sin^2\varphi \ \ \ .
\end{align}
Using $\sin^2\varphi=1-\cos^2\varphi$ and $\cos^2\varphi=\frac{1+\cos2\varphi}{2}$ the longitudinal electric field describing the longitudinal AMR results in
\begin{align}
E_{\text{x}}=\left(\frac{\rho_{||}+\rho_{\perp}}{2}+\frac{\rho_{||}-\rho_{\perp}}{2}\cos{2\varphi}\right) \, J_{\text{x}} \ \ \ . \label{eq:amr}
\end{align}
A similar consideration leads to the expression for the electric field E$_\text{y}$, transverse to the current density $J_\text{x}$. Here, the y-components of $E_{||}$ and E$_{\perp}$ have to be considered. One obtains
\begin{align}
	E_{\text{y}}&=E_{\text{y},||}+E_{\text{y},\perp}\\
		&=(\rho_{||} \, J_{\text{x}}-\rho_{\perp}\, J_{\text{x}} )\cos{\varphi} \, \sin{\varphi} \ \ \ .
\end{align}
With $\cos{\varphi}\, \sin{\varphi}=\frac{1}{2} \sin{2\varphi}$,
\begin{align}
E_{\text{y}}&=\frac{\rho_{||}-\rho{_\perp}}{2}\, \sin{2\varphi} \, J_{\text{x}} \label{eq:phe}
\end{align}
describes the transverse electric field, also called transverse AMR or planar Hall effect (PHE).

If $J_{\text{x}}$ is substituted by a thermal gradient along the x-direction ($\nabla T_{\text{x}}$), the Seebeck effect drives an electric current through the sample, leading to thermal equivalent effects of the above described current-driven effects. The Seebeck coefficient is anisotropic for the directions parallel or perpendicular to the magnetization (Fig. \ref{fig:dreieck} (b)). In case of a longitudinal measurement in an open circuit geometry the anisotropic magnetothermopower (AMTP) as the thermal counterpart of the AMR (Eq. \ref{eq:amr}) will lead to an electric field along the x-direction
\begin{align}
E_{\text{x}}= \, -\left(\frac{S_{||}+S_{\perp}}{2}+\frac{S_{||}-S_{\perp}}{2}\cos{2\varphi} \right) \, \nabla T_{\text{x}} \label{eq:amtp} \ \ \ .
\end{align}
In case of a transverse measurement the planar Nernst effect (PNE) as the thermal counterpart of the PHE (Eq. (\ref{eq:phe})) will induce an electric field along the y-direction \cite{Ky:1966b}
\begin{align}
E_{\text{y}}= \, - \frac{S_{||}-S_{\perp}}{2}\, \sin{2\varphi} \, \nabla T_{\text{x}}	\label{eq:pne} \ \ \ .
\end{align}
Since this work demonstrates the rotation of $\nabla$T in the xy-plane, a superposition of the AMTP and PNE is expected for other directions of $\nabla$T than the x- or y- axis. Assuming a measurement of the AMTP along the x- direction for rotated $\nabla$T, Eq. (\ref{eq:amtp}) adjusts to
\begin{align}
E_{\text{x}}= \, - \left(\frac{S_{||}+S_{\perp}}{2}+\frac{S_{||}-S_{\perp}}{2}\cos{2\varphi} \right) \, |\nabla T|\, \cos\varphi_T \label{eq:amtp_x} \ \ \ ,
\end{align}
with the angle $\varphi_T$ between $\nabla$T and the x-axis (Fig. \ref{fig:dreieck}). Similarly, Eq. (\ref{eq:pne}) for the PNE changes to 
\begin{align}
E_{\text{y}}= \, - \frac{S_{||}-S_{\perp}}{2}\, \sin{2\varphi} \, |\nabla T|\, \cos{\varphi_T}	\label{eq:pnex} \ \ \ .
\end{align}

Now, it has to be taken into account that the AMTP is measured along the y axis and the angles in Eq. (\ref{eq:amtp_x}) are defined with respect to the x-axis. Therefore, an angle phaseshift has to be introduced which considers that the (longitudinal) AMTP is measured along the y-axis. Keeping the angles defined along the x-direction shifts $E_{\text{y}}$ by 90$^\circ$, thus leading to
\begin{align}
E_{\text{y}}&= \, - \left(\frac{S_{||}+S_{\perp}}{2}+\frac{S_{||}-S_{\perp}}{2}\cos{\left(2(\varphi-90^\circ)\right)} \right) \, |\nabla T|\, \cos(\varphi_T-90^\circ) \nonumber \\
&= \, - \left(\frac{S_{||}+S_{\perp}}{2}-\frac{S_{||}-S_{\perp}}{2}\cos{2\varphi} \right) \, |\nabla T|\, \sin\varphi_T \ \ \ . \label{eq:amtp2}
\end{align}

Hence, Eq. (\ref{eq:amtp2}) describes the longitudinal (AMTP) and 
\begin{align}
E_{\text{y}}= \, - \frac{S_{||}-S_{\perp}}{2}\, \sin{2\varphi} \, |\nabla T| \, \cos{\varphi_{\text{T}}}	\label{eq:pne2}
\end{align}
the transverse magnetothermopower (PNE) along the y-direction. Since the experiment will measure the AMTP and PNE simultaneously, the superpositioned electric field along the y-direction is described by
\begin{align}
E_{\text{y}}&=-(S_+ \, \sin \varphi_{\text{T}} + S_- \sin (2\, \varphi - \varphi_{\text{T}})) |\nabla T|  
\end{align}
with $S_+=\frac{S_{||}+S_{\perp}}{2}$ and $S_-=\frac{S_{||}-S_{\perp}}{2}$, showing that a variation of $\varphi_{\text{T}}$ leads to a phase shift in $E_{\text{y}}$.

\section{Experimental setup}

\subsection{Applying a magnetic field}

\begin{figure}[h]
\centering
\includegraphics[width=6in]{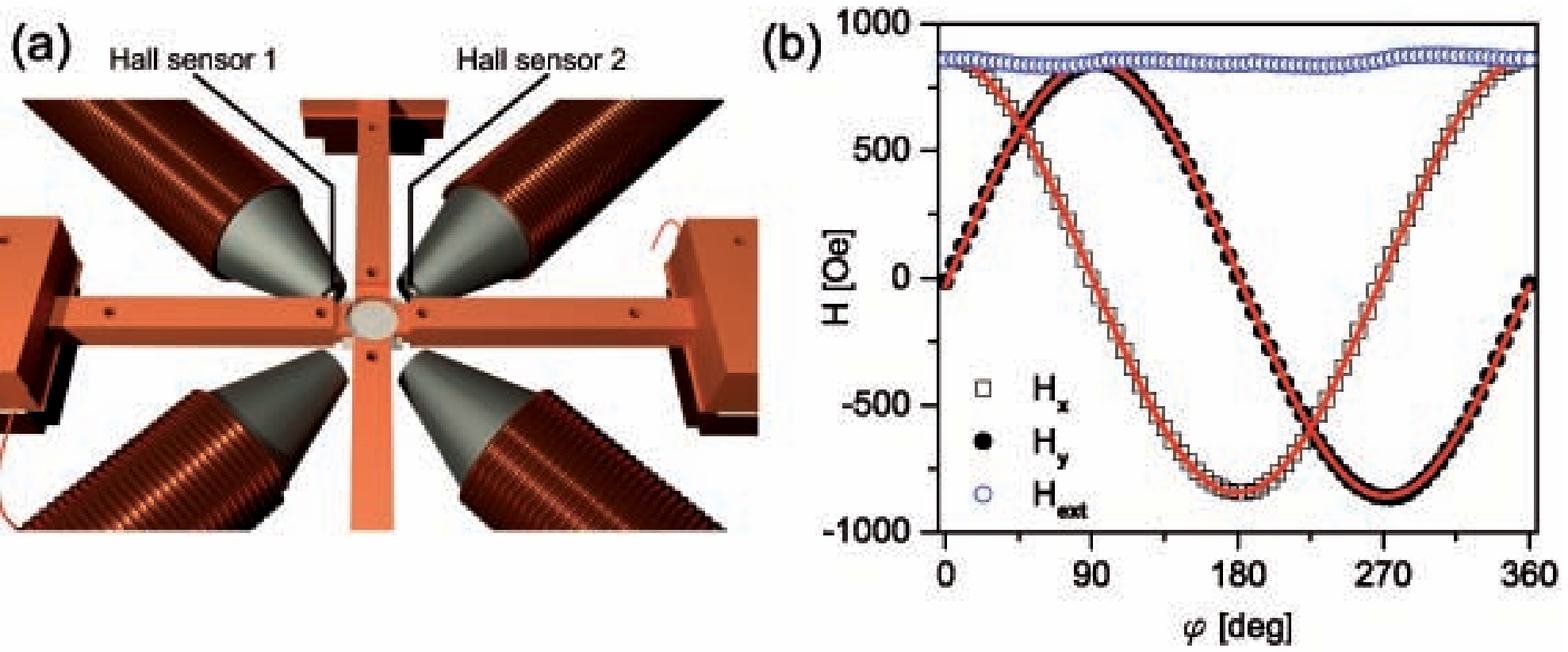}
\caption{a) Two electromagnet pairs provide the desired $\vec{H}_{\text{ext}}$ along $\varphi$ which is measured via two Hall sensors at the pole caps. b) The magnetic field rotation is verified by fitting the x- (y-) component of the applied magnetic field with a cosine (sine) function. The superposition of $\vec{H}_{\text{x}}$ and $\vec{H}_{\text{y}}$ results in $\vec{H}_{\text{ext}}$ which is constant within 1.4\%.}
\label{fig:magnet}
\end{figure}

For thermomagnetic and spin caloric measurements an external, rotatable magnetic field is eminently useful to identify the occurring transport phenomena by its symmetry. For this purpose two pairs of electromagnets with iron cores are arranged next to the sample holders (Fig. \ref{fig:magnet} a)). The distance of the iron cores can be varied, in order to be flexible for different sample sizes.

The superposition of the fields of both magnetic axes gives a net magnetic field with a selectable angle $\varphi$. Hall sensors (Projekt Elektronik, AS-NTP-Flex) are attached to the pole caps of one iron core of each magnetic axis. After calibrating its signal to the magnetic field at the sample position the simultaneous measurement of $\vec{H}_{\text{ext}}$ during the experiment is possible. In this manner magnetic fields up to $\approx$ 900 Oe can be reached. The successful magnetic field rotation is proven by measuring the x- and y-component of $\vec{H}_{\text{ext}}$ and fitting them to a cosine (sine) for $H_{\text{x}}$ ($H_{\text{y}}$) (Fig. \ref{fig:magnet} b)). The calculated resulting $\vec{H}_{\text{ext}}$ stays constant within a standard deviation of 12 Oe at a mean value of 853 Oe.

\subsection{Electric contacting}
The setup allows the installation of four micro-probe measurement stages to electricly contact the samples via, e.g., contact needles. For the lowest background noise (RMS $\approx \, 50\,$nV) the samples were bonded with 25 $\mu m$ thin Au wires whose ends were glued to Au probes by silver paste. The electric signal is then detected by a nanovoltmeter (Keithley 2182A).

\subsection{Detecting the temperature profiles}
\begin{figure}[h]
\includegraphics[width=3.3in]{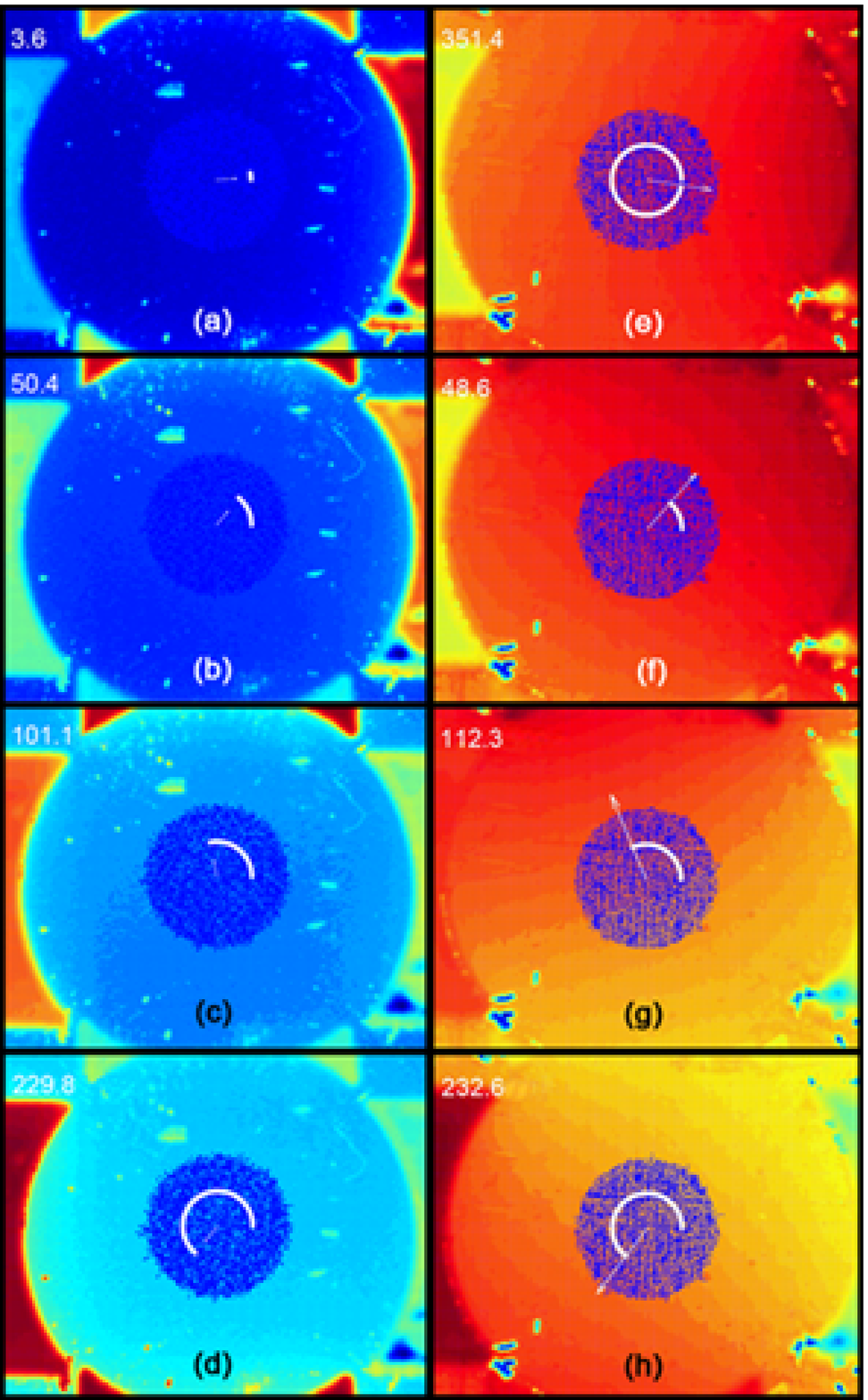}
\caption{Thermographic pictures taken for (a)-(d) MgO and (e)-(h) Cu for (a), (e) $\varphi_{\text{T}}=0^\circ$; (b), (f) $\varphi_{\text{T}}=45^\circ$; (c), (g) $\varphi_{\text{T}}=105^\circ$; (d), (h) $\varphi_{\text{T}}=225^\circ$. Small deviations between the applied $\varphi_{\text{T}}$ and the calculated average angle of up to 6$^\circ$ can be detected due to surface defects, inhomogenous $\nabla$T, different thermal conductivities or a relative rotation between the setup and the infrared camera.}
\label{fig:irrotT}
\end{figure}

In addition to the electric temperature detection, a FLIR SC7000 infrared camera is used to detect the temperature profiles of the sample. The vertical mount of the IR camera 0.5$\,$m above the sample allows the optical detection of the heat distribution directly at the sample surface. In contrast to the electric temperature detection by the PT1000 thermometers this method excludes any temperature losses via heat conduction, e.g., due to the Cu clamps. These data are used to quantify parameters like the magnitude of temperature gradient, base temperature and angle of the applied $\nabla$T as follows: First, a circular region of interest (ROI) is defined where data should be considered from for the following calculations and which excludes possible surface defects (blue circles in Fig. \ref{fig:irrotT}). Second, for each data point in the ROI the local temperature gradient in x- and y- direction is calculated which leads to a determination of an average temperature gradient. These quantities are used to identify the average angle under which $\nabla T$ is applied (symbolized by the white arcs in Fig. \ref{fig:irrotT}). Third, the vectorial temperature profile along the calculated angle is used for a linear fit whose origin is the temperature at the center of the ROI (base temperature of sample) and whose slope is equal to $\nabla$T.

As a first proof of the successful rotation of $\nabla$T, copper- (Cu-) and MgO- substrates were thermographically investigated. 
Fig. \ref{fig:irrotT} shows pictures taken from the infrared camera for a MgO and Cu substrate, after applying $\nabla$T at $\varphi_T=45^\circ, 105 ^\circ, 225^\circ \text{ and } 360^\circ$. Although a deviation between the applied $\varphi_{\text{T}}$ and the calculated average output angle is observed, the rotation of $\nabla$T could be controlled in this measurement within an uncertainty of 6$^\circ$. 

\section{Superposition of magnetic anisotropies}
\begin{figure}[h]
\includegraphics[width=4in]{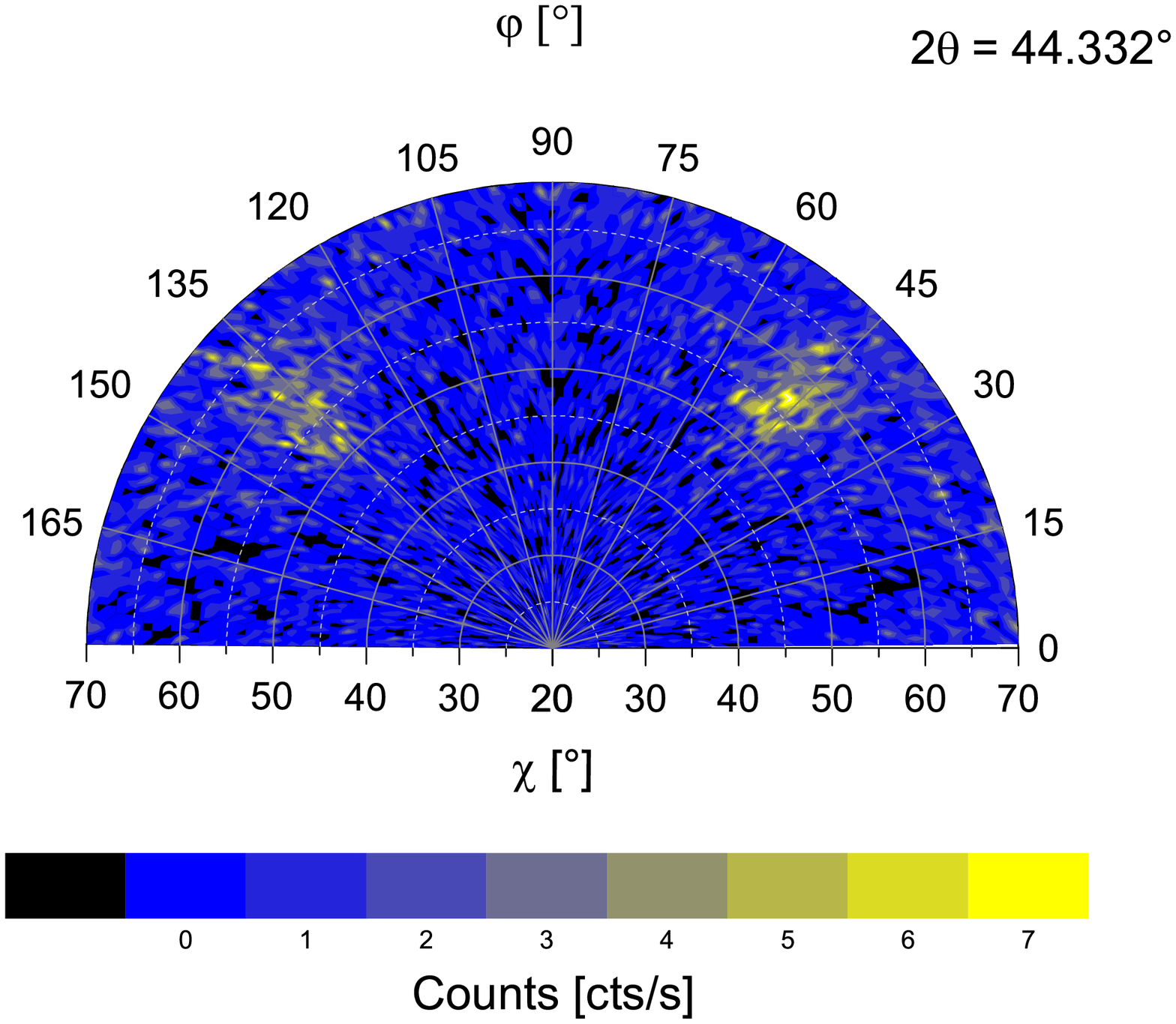}
\caption{XRD measurement of Py/MgO(001) at 2$\Theta=44.332^\circ$ shows a fourfold diffraction pattern, indicating a cubic crystal structure.}
\label{fig:euler}
\end{figure}

The appearance of two magnetic easy axes (MEA, $\varphi=20^\circ,\, 70^\circ$) can be explained by the superposition of a uniaxial (UMA) and a fourfold in-plane cubic magnetic anisotropy (CMA). A UMA can be attributed to, e.g., surface steps\cite{Chen:1992}, oblique growth \cite{Park:1995}, substrate shape \cite{Kuschel:2011} or dangling bonds\cite{Daboo:1995}. For example, Fe/MgO(001) systems were used to manipulate the UMA in terms of strength and orientation with different deposition techniques \cite{Zhan:2007,Zhan:2009} showing the likelihood of forming a UMA during the deposition of thin films, not only for Fe/MgO(001) but also for permalloy thin films on different substrates \cite{Kaibi:2015, Li:2015}. Due to the crystalline symmetry of cubic magnetic films a CMA is expected to be present for cubic systems \cite{Zhan:2007, Zhan:2009}. X-ray diffraction (XRD) measurements on our Py/MgO(001) sample show a fourfold diffraction pattern at a 2$\Theta$-angle of 44.332$^\circ$ for (111) Bragg reflections (Fig. \ref{fig:euler}), which can clearly be separated from the MgO(001) signal. Therefore, the presence of a CMA is very likely. 




The magnetic field dependent voltages in Fig. 4 and 7 in III. B and III. C (main paper) have partially asymmetric behavior, although AMTP/PNE traces are known to be symmetric \cite{tsse2,prlmax,AMTEP,AMTEP2,AMTEP3,timo3} to the magnetic field in case of a present UMA. To resolve this situation different simulations were conducted and described in the following.
For Eqs. (15), (16) it is implicitly assumed that the magnetization vector $\vec{M}$ coincides with the magnetic field $\vec{H}$ which means that $\varphi_{M0}=\varphi$, with $\varphi_{M0}$ as the angle of equilibrium position of $\vec{M}$. This assumption works well when the applied magnetic field is at least one order of magnitude stronger than the magnetic anisotropy or any parasitic magnetic field $\vec{H_p}$ contribution in the system \cite{gurevich}: $\left | H \right |>> \frac{C_i K_i}{M_s}, \left | H_p \right |$, where $K_i$ is some anisotropy constant with the dimension of J/cm$^3$, $M_s$ is the saturation magnetization and $C_i$ is a dimensionless constant of the order of unity (depending on the type of anisotropy). However, this assumption is fulfilled only for external magnetic fields greater than 100 Oe. Thus, an adequate model for calculations of $\varphi_{M0}$ for lower external magnetic fields is necessary, in order to reproduce complete field sweeps of the experimental AMTP/PNE traces. 

First of all, using $V_y=-E_y d$ (\emph{d} is the distance between measuring contacts) Eq. (17) can be rewritten as
\begin{align}
V_{y}(H)=d \, S_- \left | \nabla T \right |\sin(2\varphi_{M0}(H)-\varphi_{T})  \label{V_ysim} \ \ \ ,
\end{align}

where the summand with $S_{+}=\frac{S_{\parallel}+S_{\perp}}{2}$ is neglected, since it does not depend on $\vec{H}$ and produces only an offset in the magnetic field sweep measurements. 
In further calculations and in Figs. 5 and 8 of the main paper normalized AMTP/PNE traces are shown ($d \, S_{-}\left | \nabla T \right |=1$). The experimental data of magnetic field sweeps (Fig. 4, main paper) hint to the presence of a magnetic anisotropy. Magnetic easy directions (magnetic easy axes MEA1 and MEA2) are situated around $\varphi_{min1}=$ 20$^\circ$ and $\varphi_{min2}=$ 70$^\circ$, since the signal amplitudes are smallest in these directions. This kind of anisotropy could result from the combination of UMA and CMA of the sample which is not parallelly aligned as mentioned previously \cite{Kuschel:2012, Zhan:2007, Zhan:2009, UMA-CMA2}.

\begin{figure}
	\centering
		\includegraphics[width=10cm]{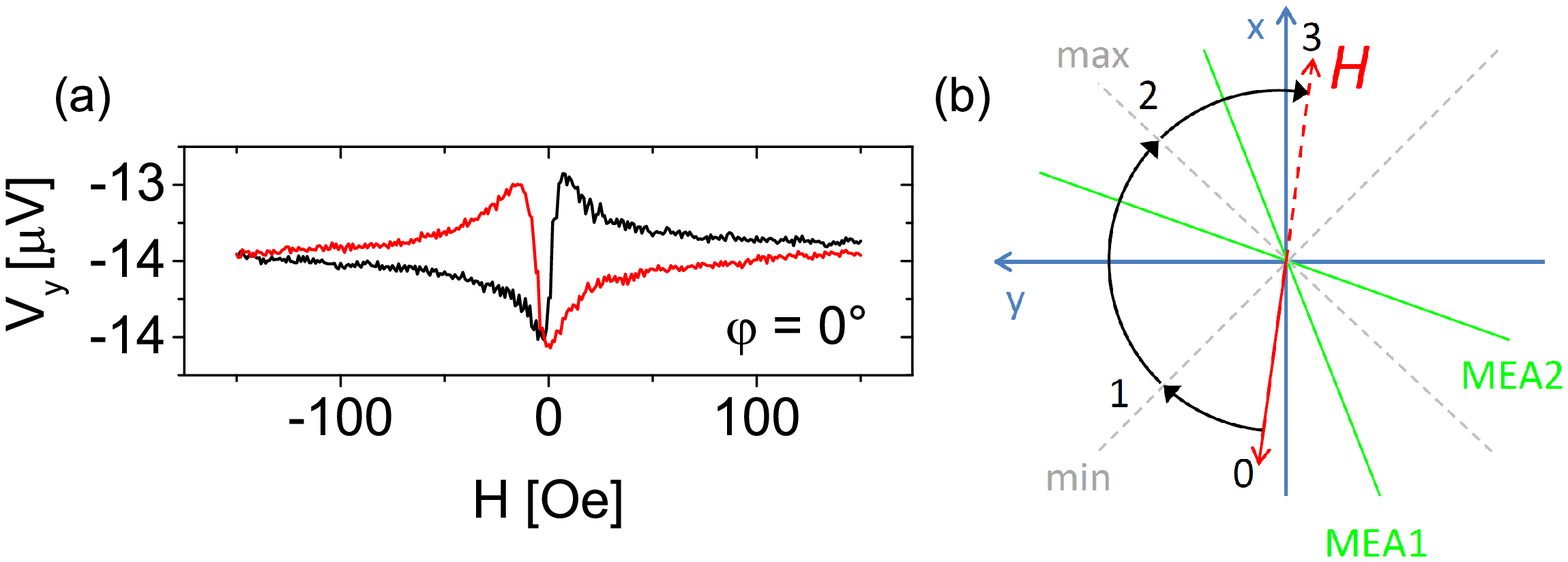}
		\caption{Explanation of experimental field sweeps: (a) The experimental sweep up trace for $\varphi=0^\circ$ passes through 4 specific positions, (b) $\vec{M}$ reversal in relation to the experiment at $\varphi=0^\circ$. MEA1 and MEA2 (green lines) are magnetic easy axes situated around $20^\circ$ and $70^\circ$, respectively.}
		\label{1sweep}
\end{figure}

As a first suggestion for the rotation of $\vec{M}$, but later proven by model C, the shape of the experimental curves can be understood with the help of Fig. \ref{1sweep} for the sweep up curve at $\varphi=0^\circ$ and $\varphi_T=0^\circ$. According to Eq. (\ref{V_ysim}) the maximum and minimum of the signal appears for $\varphi_{M0}=45^\circ,\, 225^\circ$ and $\varphi_{M0}=\, 135^\circ,\, 315^\circ$ respectively (dashed, gray lines in Fig. \ref{1sweep}(b)). Point \bfseries{0} \normalfont is the fully saturated state of the sample, when $\vec{M}$ is aligned with the field $\vec{H}$ (corresponding to negative values of $\vec{H}$). When the absolute field value of $\vec{H}$ goes down, $\vec{M}$ tends to rotate off the direction of the external magnetic field and $\vec{M}$ passes through the direction of minimum \bfseries{1} \normalfont (despite it is expected to rotate in the opposite direction, closer towards MEA1). When the magnetic field switches its direction (towards red dashed arrow) and increases its absolute value, $\vec{M}$ passes through the maximum position \bfseries{2} \normalfont trying to align with the magnetic field in direction \bfseries{3}\normalfont.

\begin{description}
\item[Model A] According to Gurevich \emph{et al.}\cite{gurevich}, the in-plane density of magnetic free energy $U$ in the presence of a UMA and CMA for monodomain magnetization reads
\begin{align}
U=-M_{S}H\cos(\varphi_{M}-\varphi)+K_U\sin^{2}(\varphi_{M}-\varphi_{UA})+\frac{K_C}{4}\sin^{2}2(\varphi_{M}-\varphi_{CA}) \ \ \ ,
\label{energ}
\end{align}
where the first term is the Zeeman energy, the second and third terms represent the magnetocrystalline uniaxial and cubic anisotropy energies, $M_{S}$ is the saturation magnetization,  $K_{U}$ and $K_{C}$ are the constants describing the strength of the UMA and CMA with angles $\varphi_{UA}$ and $\varphi_{CA}$ respectively, and $\varphi_{M}$ is an arbitrary direction of the magnetization vector $\vec{M}$.
The demagnetization energy is excluded, since its in-plane contribution for the given geometry is negligible. The demagnetization factors are calculated according to Aharoni \emph{et al.} \cite{demag} which leads to an effective in-plane demagnetization factor $(N_x-N_y)/4\pi$ of the order of $10^{-6}$ (effective in-plane demagnetizing field $\approx 0.01$ Oe). Using reasonable values of $K_{C}=5\cdot 10^{4}$ erg/cm$^3$ \, \cite{CMAK}, $K_{U}=2\cdot 10^{4}$ erg/cm$^3$ \, \cite{UMAK, pypaper}, $\varphi_{CA}=0^\circ$ and $\varphi_{UA}=45^\circ$ and without an external magnetic field, the free magnetic energy angular distribution shown in Fig. \ref{energyfig} is calculated. As can be seen, the easy axes are aligned to the previously mentioned directions 20$^\circ$ and 70$^\circ$. However, with application of this model \cite{pypaper} only symmetric PNE traces can be obtained, because $\vec{M}$ switches between directions 1 and 2 immediately via multidomain state after inverting $\vec{H}$ (described in Fig. \ref{mumax}(b)). A multidomain state is only supposed to exist in an extremely narrow range around $\vec{H}$ = 0 Oe, which is a reasonable assumption for Py thin films.
\begin{figure}[h]
	\centering
		\includegraphics[width=10cm]{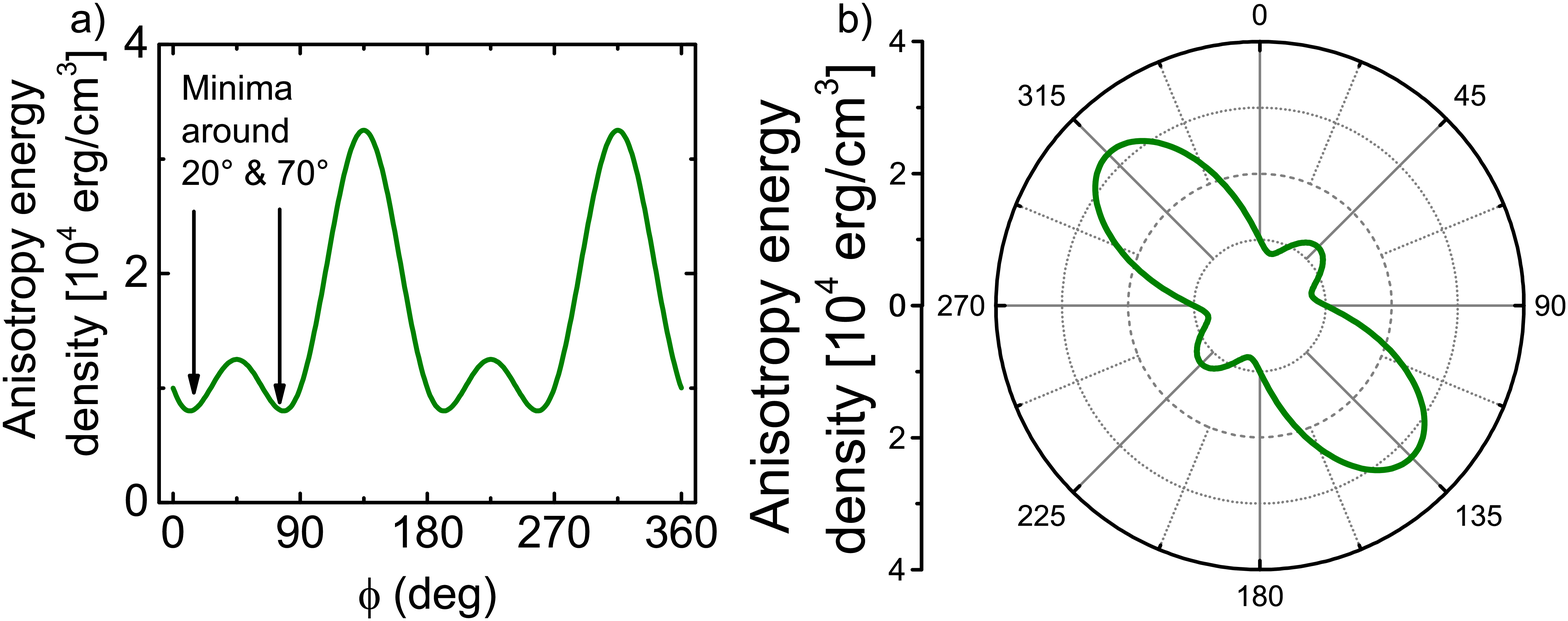}
		\caption{(a) Energy angular dependence according to Eq. (\ref{energ}) with parameters $K_{C}=5\cdot 10^{4}$ erg/cm$^3$, $K_{U}=2\cdot 10^{4}$ erg/cm$^3$, $\varphi_{CA}=0^\circ$ and $\varphi_{UA}=45^\circ$. Magnetic easy directions are situated around $\varphi_{min1}=$ 20$^\circ$ and $\varphi_{min2}=$ 70$^\circ$;
		(b) Representation of energy landscape in polar coordinate system.}
		\label{energyfig}
\end{figure}

\item[Model B] A more elaborate investigation with keeping the magnetic prehistory of the sample (hysteresis phenomenon), using multidomain state and including modeling of finite temperatures as well as UMA and CMA was made in MuMax3  \cite{mumax3}, leading to Fig. \ref{mumax}. In this model it is not possible to conduct simulations for large sample areas (5x5 mm$^2$), because of the cell number limitation of MuMax3. Instead, an area of 1x1 $\mu $m was used. The reduction of area leads to an enhanced contribution of demagnetization energy to anisotropy by 3 orders of magnitude. To make this contribution negligible, $M_S$ has to be reduced twice in comparison to permalloy's value and the anisotropy constants were enlarged by an order of magnitude in comparison to model A. The sweep region was expanded to [-1000; 1000] Oe. Thus, this model only gives a qualitative description. The simulated voltage trace in Fig. \ref{mumax} (a) gives the antisymmetric behaviour, but yet the behavior of $\vec{M}$ in Fig. \ref{mumax}(b) is very different to the one shown in Fig. \ref{1sweep}(b). $\vec{M}$ rotates in direction \bfseries 1 \normalfont which corresponds to MEA1, when the field sweep goes up. Then $\vec{M}$ switches its direction by $180^\circ$ via multidomain state (the region between \bfseries 1 \normalfont and \bfseries 2 \normalfont in Fig. 7 (a); position 1' is related to $H\approx 0$ Oe) into \bfseries 2 \normalfont. The minimum of the voltage trace does not occur because of passing through minimum direction at $135^\circ$, but because of the multidomain state, which tends to reduce the absolute value of $\vec{M}$ (AMTP/PNE signal is $\propto \left | M \right |^2$). Next, when $\vec{H}$ switches its direction and increases its value, $\vec{M}$ tends to align with the external magnetic field (point \bfseries  3\normalfont). Asymmetric behaviour in systems with CMA, low $M_S$ and large magnetic anisotropy was already observed in other AMTP/PNE experiments \cite{AMTEP2,timo3}.
\begin{figure}[h]
	\centering
		\includegraphics[width=10cm]{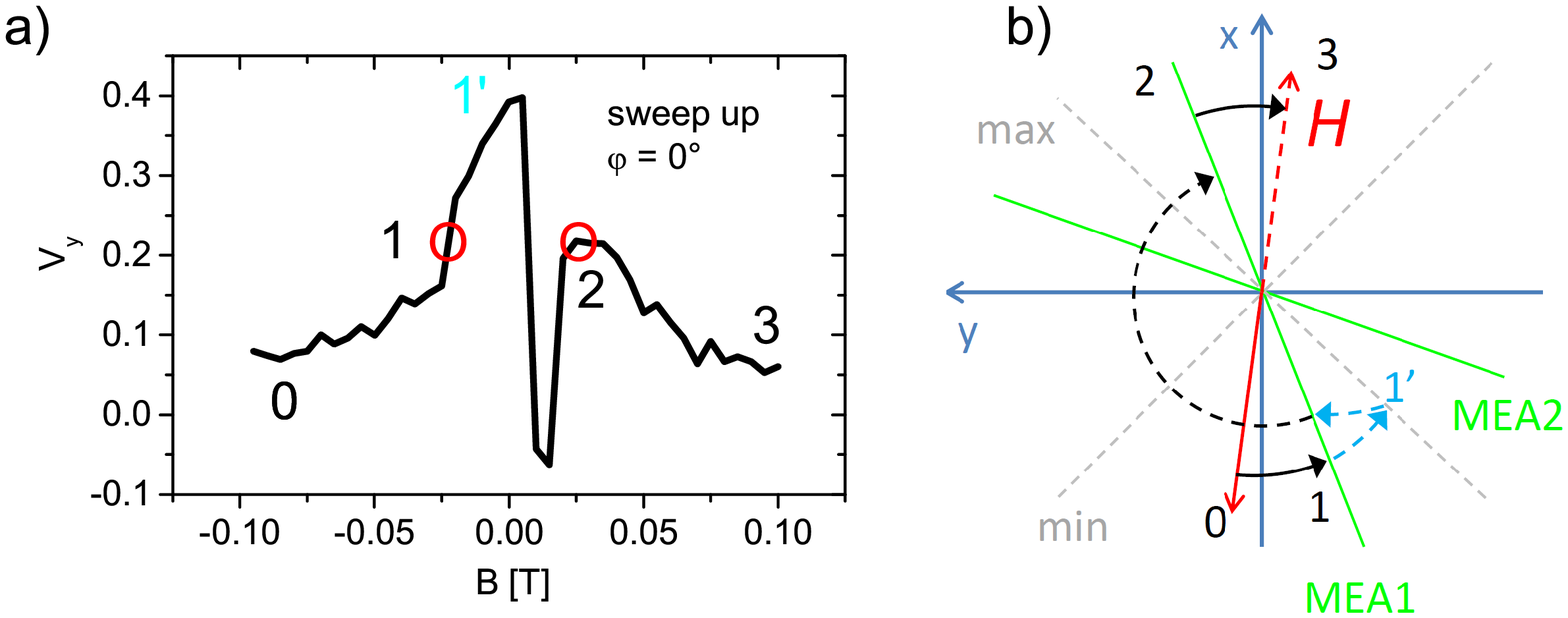}
		\caption{MuMax3 simulation for sweep up curve: (a) normalized voltage sweep up curve for $\varphi=0^\circ$; (b) $\vec{M}$ behavior during field sweep up for $\varphi=0^\circ$.} 
		\label{mumax}
\end{figure}

\item[Model C] Next, the directions of UMA and CMA were analyzed in more detail. The fact that the magnetic hard directions of the CMA coincide with the directions of the magnetic pole caps of the setup, hints to a magnetic anisotropy which rather comes from the magnet system itself than from the sample. In case of a perfect geometry of the magnet a pure CMA could be expected only due to its symmetry. But with slightly different distances between the opposing magnetic cores and inhomogeneities in the yoke an additional introduction of UMA seems to be possible, just because one magnetic pair might be more preferable for the magnetic flux than the other one. Magnetic hard directions of CMA coincide with the magnetic poles of the magnet, since even when only one pair is used, there is always some remanence magnetization in the second pair. Thus, the fields of both pairs always have to be summed up, so that the superpositioned field direction always tends to stay somewhere between the pole directions.

Now, a parasitic field $H_p$ due to the magnetic yoke is assumed. Because of the direct connection of both magnetic pairs through the toroidal yoke it may lead to a "leakage" of magnetic flux from one pair to the other: The magnetic state of one pair is sensitive to the state of the second pair, while the permeability of the yoke has nonlinear dependence on the magnetic flux passing through it. The magnetic flux of one pair modulates the magnetic transparency of the yoke and influences the behavior of the second pair of poles. With these considerations, the parasitic magnetic field $H_p(\varphi)$ can be written as
\begin{align}
\left | H_p(\varphi) \right |=\left | H_{pmax} \right |\left |\sin(\varphi-\varphi_{min1})\sin(\varphi-\varphi_{min2}) \right | \ \ \ ,
\label{parasabs}
\end{align}
with $H_{pmax}=7.5$ Oe as the modulus of parasitic field, $\varphi_{min1}=20^\circ$ and $\varphi_{min2}=70^\circ$. $H_p(\varphi)$ is orientated along
\begin{align}
\varphi_p(H, \varphi)=\varphi\pm180^o \left(\frac{H_{max}+H}{2H_{max}}\right ) \ \ \ ,
\label{parasangle}
\end{align}
with $H_{max}$ as the amplitude of the applied magnetic field during the sweep (150 Oe). $\left | H_p(\varphi) \right |$ (see simulated trace Fig. \ref{Hp}) takes into account the anisotropic character of the magnet system (amplitude of anisotropy). Its angular dependence behaves very similar to energy angular dependence of model A shown in Fig. \ref{energyfig}. The phase shift (second summand of angle of parasitic field $\varphi_p$ in Eq. (\ref{parasangle})) takes the "leakage process" and anisotropic character (rotation of $\vec{M}$) into account as well. The sign of the phase shift ($+$ or $-$) is chosen the way that $\vec{M}$ rotates in the direction of the closest minimum of AMTP/PNE signal (Fig. \ref{1sweep}(b)). $\varphi_{CA}=0^\circ$ of CMA is chosen according to XRD data from Fig. \ref{fig:euler} and $K_{C}$ is the same as in model A.

\begin{figure}[h]
	\centering
		\includegraphics[width=5cm]{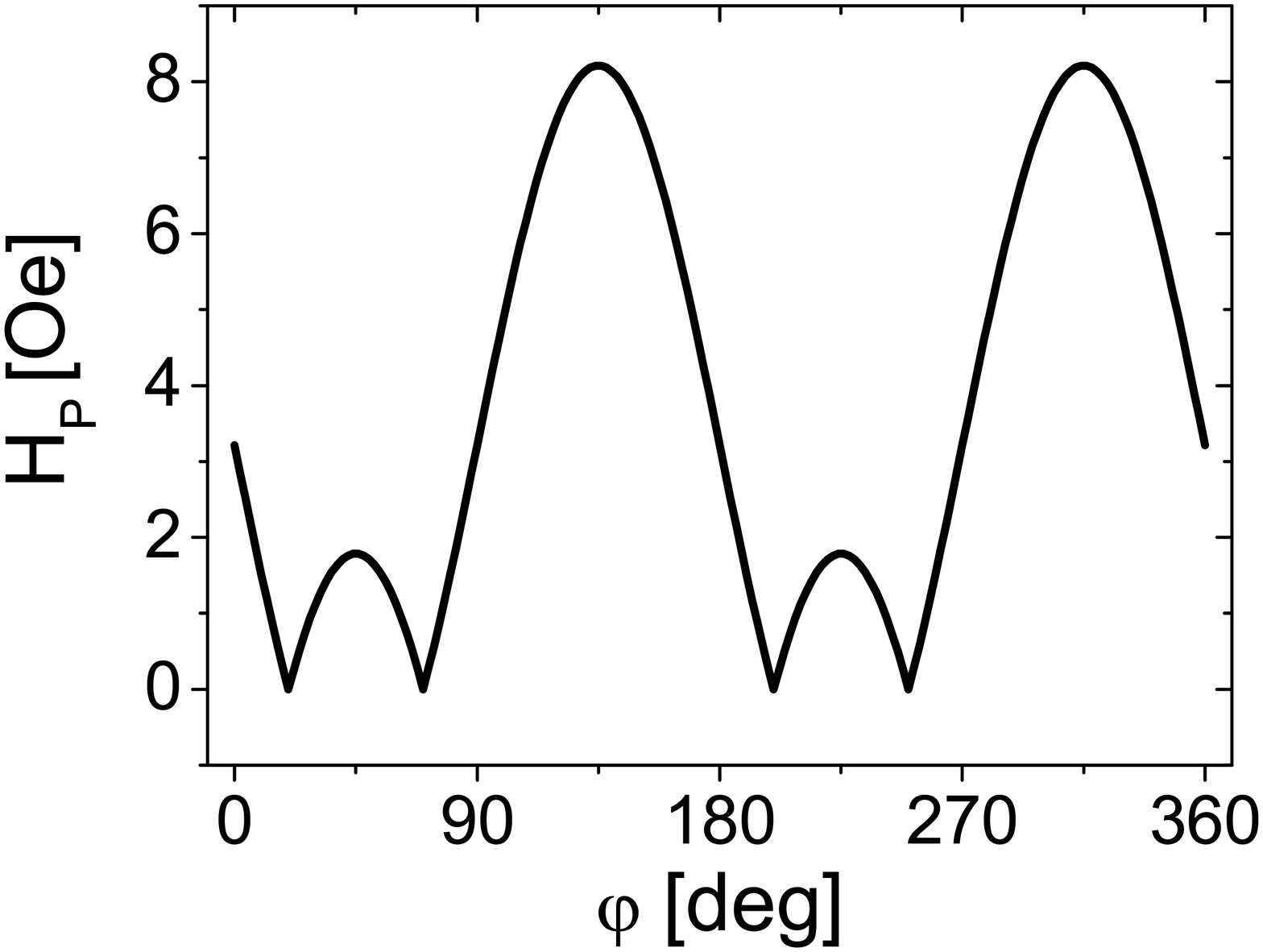}
		\caption{ $\left | H_p(\varphi) \right |$ according to Eq. (\ref{parasabs}) with $H_{pmax}=7.5$ Oe, $\varphi_{min1}=20^\circ$ and $\varphi_{min2}=70^\circ$.} 
		\label{Hp}
\end{figure}

 The subsequent calculation for each sweep measurement at a specific angle $\varphi$ includes the following steps:

\begin{enumerate}
\item The parameters $H_{pmax}, \varphi_{T}, \varphi_{min1}, \varphi_{min2}$ are kept fixed. First, $\left | H_p(\varphi) \right |$ is calculated. For each value of the sweep field ${H}$, $\varphi_p(H)$ is determined;
\item In Eq. (\ref{energ}) $H$ is substituted with $H_{\Sigma}(H)$ (effective magnetic field) and $\varphi$ with $\varphi_{\Sigma}(H)$ (effective angle). $K_{c}=5\cdot 10^{4}$ erg/cm$^3$, $K_{u}=0$ erg/cm$^3$, $\varphi_{ca}=0^\circ$ are used. In order to find $\varphi_{M0}(H)$ the derivative of Eq. (\ref{energ}) is taken and its root is determined (similar to Shestakov \emph{et al.} \cite{pypaper}). This procedure is done for each value of $H$. After this step the $\varphi_{M0}(H)$ dependence is known.
\item After inserting the computed dependence $\varphi_{M0}(H)$ into Eq. (\ref{V_ysim}) the normalized curve $V_{y}(\varphi_{\Sigma}(H))$ (with $d S_{-}\left | \nabla T \right |=1$) can be constructed and compared with the experimentally determined data.
\end{enumerate}

\end{description}

The calculated traces (Fig. 5 and Fig. 8, see main manuscript) fit the experimental data (Fig. 4 and Fig. 7, see main manuscript) qualitatively well, despite the use of a relatively simple model.